\documentclass[journal]{IEEEtran}

\usepackage{graphics}
\usepackage{amsmath}
\usepackage{amssymb}
\usepackage{amsfonts} 	    % used for math \in symbol

\usepackage[utf8]{inputenc} % for ü in Zürich
\usepackage[inline]{enumitem}

\usepackage[algo2e]{algorithm2e}
\usepackage{algorithm}
\usepackage{mathtools}

\usepackage{enumitem}

\usepackage{ wasysym } % for cent symbol

\usepackage{hyperref}

\graphicspath{ {images/} }

\usepackage{tikz} % for plotting
\usepackage{csvsimple}
\usepackage{pgfplots}
\pgfplotsset{compat=1.8}
\usepackage{pgfplotstable}
\usepgfplotslibrary{groupplots, dateplot}

\definecolor{matlabblue}{RGB}{0,114,189}
\definecolor{matlabyellow}{RGB}{238, 177, 32}
\definecolor{matlabpurple}{RGB}{126, 47, 142}
\definecolor{matlabred}{RGB}{217, 83, 25}

\DeclareMathOperator{\range}{range}

\usepackage{siunitx}
\allowdisplaybreaks

\title{\LARGE \bf
Multiperiod Stochastic Peak Shaving Using Storage
}

\author{Benjamin Flamm, Guillermo Ramos, Annika Eichler, John Lygeros % <-this % stops a space
\thanks{The authors are with the Automatic Control Laboratory of ETH Zürich, Physikstrasse 3, 8092 Zürich, Switzerland.
        {\tt\small \{flammb, eichlean, lygeros\}@control.ee.ethz.ch}, {\tt\small w.ramosdb@gmail.com}}%
}

\begin{document}

\maketitle

\begin{abstract}
We present an online stochastic model predictive control framework for demand charge management for a grid-connected consumer with attached electrical energy storage. The consumer we consider must satisfy an inflexible but stochastic electricity demand, and also receives a stochastic electricity inflow. The optimization problem formulated solves a stochastic cost minimization problem, with given weather forecast scenarios converted into forecast demand and inflow. We introduce a novel weighting scheme to account for cases where the optimization horizon spans multiple demand charge periods. The optimization scheme is tested in a setting with building demand and photovoltaic array inflow data from a real office building. The simulation study allows us to compare various design and modeling alternatives, ultimately proposing a policy based on causal affine decision rules.
\end{abstract}

\section{Introduction}
The electricity bill for large industrial electricity consumers often includes a portion that penalizes the customer's maximum demand over a given period, e.g., each month. This demand charge recoups investments made in the transmission and distribution infrastructure, since the electricity grid must be sized for the maximum load encountered. As this charge can be a significant portion of the total electricity bill, e.g., 40\% for the case study in \cite{Glassmire2012}, it is advantageous for large consumers to engage in demand charge management (DCM), where the grid-facing electricity demand is reduced during periods where a peak would otherwise be expected. 

This approach, also known as peak shaving, can function in several ways. First, consumers can use flexibility built into energy-consuming processes to temporally shift consumption. Second, consumers can store energy in order to spread the high peak energy usage over time. As \cite{Neubauer2015} notes, the effect of DCM is dependent on the load profile and how the profile overlaps with associated power generation. 

Optimal control schemes for storage include various energy-related components in their cost functions. \cite{Ma2011} considers both time-of-use energy and demand charges in a model predictive control (MPC) setting, but the demand charge is only accrued for the current day. Since peaks from common industrial loads like office buildings follow a roughly daily pattern, considering only a day horizon can be a good approximation of the monthly peak reduction problem \cite{Hanna2014}. \cite{Narimani2017} considers the sum of the maximum monthly demand as well as an additional penalty on the maximum demand during certain times of day.

A real-time control scheme must also account for the stochasticity inherent in forecasts. Some papers use a single forecast, such as \cite{Hanna2014}, which uses the load realization from the previous day as a forecast for the coming day. Others use statistical methods. \cite{Conte2018} provides a mixed-integer stochastic optimization method based on chance constraints to maximize an economic objective, with uncertainty in photovoltaic (PV) production. \cite{Yu2017} also considers the stochastic optimization of storage subject to demand and energy charges, deriving structural results based on perfect efficiency and sufficient storage size. \cite{Jin2017} models the stochasticity using an exogeneous Markov chain, and incorporates this directly into a dynamic program.

After choosing models for the objective function and stochasticity, the next task is to formulate the  problem and choose a solution method. Some papers forego optimization altogether and consider heuristic policies. For example, \cite{Park2017} proposes a policy that charges or discharges based on the net electricity demand. Optimization-based schemes that consider DCM often include a state denoting the previous maximum demand observed in the given period \cite{Ma2011,Jin2017,Jones2017}. The latter two papers then solve the problem as a dynamic program. 

In this paper, we seek to minimize the cost of meeting an electricity demand using a grid-connected generic battery energy storage system (BESS) in an MPC fashion, subject to charges on the electricity grid usage. We formulate the DCM problem as a linear program (LP), as is done in \cite{Hanna2014}. Our work extends earlier results in the literature in several directions. First, we solve the problem in an MPC setting, minimizing operating costs over a receding horizon that can overlap with multiple demand periods. Second, we propose a novel weighting for energy and peak costs over prediction horizons shorter than a peak period. Third, we forecast PV production and building demand based on real weather forecasts and data, and update these models in an online manner to reduce prediction error. Finally, we apply a causal online policy to a real system, achieving performance within 1.3\% of the optimal result when all data is known.

In Section \ref{section:models_and_data}, we present the system data and weather forecasts considered. In Section \ref{section:pv_inflow_model}, we formulate a model to predict PV output based on weather forecasts. In Section \ref{section:demand_model}, we do the same for building demand. Section \ref{section:deterministic_problem} formulates the best-case deterministic problem, while Section \ref{section:stochastic_problem} modifies the setting to include stochastic forecasts. Finally, Section \ref{section:results} presents simulation results for the deterministic and stochastic settings.

\section{Forecast and System Data} \label{section:models_and_data}
In the subsequent sections, we develop predictive models for PV production and building demand, based on provided weather forecasts. We first present the data to be modeled.

\subsection{PV inflow and electricity demand data}
\begin{figure}
    \centering
 \begin{tikzpicture}
  \begin{groupplot}[
     group style = {group size = 1 by 2},
     height = 4 cm,
     width = 0.99\linewidth,
     grid=major,
     xmin=2018-03-01, 
     xmax=2018-05-04,
     ymin=0, 
     ymax=100
    ]
    \nextgroupplot[date coordinates in=x, table/col sep=comma, xticklabel=\year-\month,xtick = {2018-03-01, 2018-04-01, 2018-05-01},
  legend style={at={(0.2,0.97)}, anchor=north},
  ylabel={PV [\SI{}{\kilo \watt}]}
  ]
   \addplot[no markers,matlabblue] table[x = date_array, y = pv] {deterministic_data_2018_mod.csv};
   
   \nextgroupplot[xlabel=Date, ylabel={[\SI{}{\kilo \watt}]}, date coordinates in=x, table/col sep=comma, xticklabel=\year-\month,xtick = {2018-03-01, 2018-04-01, 2018-05-01}, 
  legend style={at={(0.2,0.18)}, anchor=north},
  ylabel={Demand [\SI{}{\kilo \watt}]},
  ]
   \addplot[no markers,matlabblue] table[x = date_array, y = demand] {deterministic_data_2018_mod.csv};
  \end{groupplot}
\end{tikzpicture}
    \caption{PV production and electricity demand for PSI PV array and WBWA office building from Mar. 1 - Apr. 30, 2018.}
    \label{fig:pv_demand_data}
\end{figure}
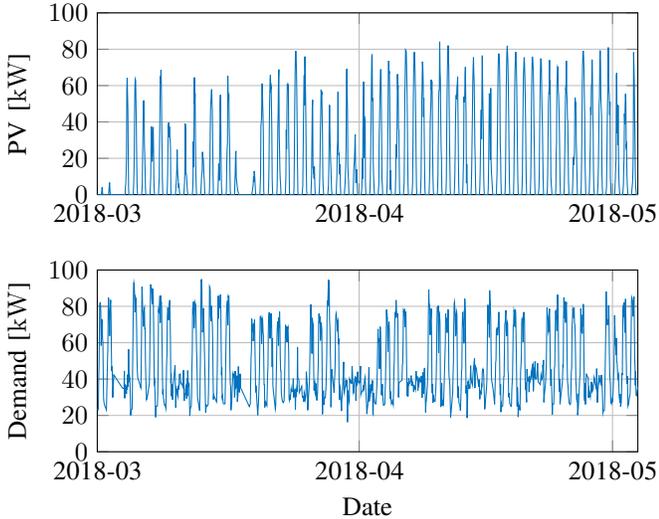

We consider the power output of a PV array located at the Paul Scherrer Institute (PSI) in Villigen, Switzerland. The data, which has a sample rate that varies from several seconds to several minutes per sample, is averaged over hourly intervals. As shown in the first graph in Fig.~\ref{fig:pv_demand_data}, PV production follows a diurnal cycle, with a time average of \SI{16.5}{\kilo \watt} and a peak value of \SI{84.1}{\kilo \watt}.

We also consider the hourly electricity demand data from an office building on the PSI campus. As shown in the second graph in Fig.~\ref{fig:pv_demand_data}, the demand also follows a roughly diurnal load, with a mean value of \SI{46.7}{\kilo \watt}, and a peak value of \SI{94.7}{\kilo \watt}. The graph also shows that weekend and holiday demand has a different shape than on weekdays.

\subsection{Weather forecasts}
We use the COSMO-E five-day-ahead weather forecast provided by the Swiss Federal Office of Meteorology (MeteoSwiss) \cite{MeteoswissLink}. The forecasts consist of an ensemble of 21 different predictions of meteorological variables, with hourly data points for the next five days (120 hours). New forecasts are generated every 12 hours, at midnight and noon. The forecasts are given over a grid of points covering Switzerland, with a grid spacing of 2.2 km. In the subsequent modeling, we use the forecasts of air temperature at 2 meters above ground level and downward shortwave radiation flux at the surface (in \SI{}{\watt \per \meter^2}), chosen at the gridpoint nearest to the PSI campus in Villigen, Switzerland (47.54$^{\circ}$N 8.23$^{\circ}$E).

\section{PV Inflow Model and Weather Classification Algorithm} \label{section:pv_inflow_model}
We wish to predict the production of the PV array presented above, as a function of the COSMO-E forecasts of solar irradiance and air temperature. We train separate prediction models for clear and cloudy days, as done in \cite{Chen2011}. Each day of each weather forecast scenario is classified as clear or cloudy using the algorithm described in Section 3.2 of \cite{Reno2012}. We now compare two candidate models: one model with a small number of parameters and one based on an artificial neural network (ANN).

\subsection{Regression-based model for PV prediction}
As a simple candidate model, we choose a version of the well-known PVUSA model \cite{Dows1995}. This assumes the generated PV power $P^\mathrm{PV}$ can be expressed as a function of the solar irradiance and air temperature in the following manner:
\begin{equation} \label{eq:pvusa_output}
    P^\mathrm{PV} = \gamma_1 I + \gamma_2 I^2 + \gamma_3 IT
\end{equation}
where $I$ is the solar irradiance, $T$ the air temperature, and $\gamma_1$, $\gamma_2$, and $\gamma_3 \in \mathbb{R}$ are the model parameters. We determine the parameters based on a least-squares fit of historical forecasts and the corresponding PV output realization.

We fit one model each for clear and cloudy forecasts, and only use the one-day-ahead forecasts for training. The reasoning is that the forecast accuracy decreases as the horizon increases, affecting the prediction performance of the regression-based models.

\subsection{Artificial neural network for PV prediction} \label{section:pv_ann_architecture}
In a similar manner to the regression-based model, we train artificial neural networks (ANNs) on the PV data separately for cloudy and clear forecasts. We choose the following ANN structure based on cross-validation on historical PV data:
\begin{enumerate*}[label=(\roman*)]
\item an input layer of width 48, corresponding to a scenario of 24 solar irradiance and 24 air temperature forecasts,
\item a fully connected layer of width 96, with a dropout layer with probability 25\%, and a rectified linear unit (ReLU) activation function, and
\item an output layer of width 24, corresponding to the next 24 values of PV production.
\end{enumerate*}

We normalize the training data to the interval $[0, 1]$. The network is trained for 500 epochs using the ADAM optimizer \cite{ADAMoptimizer}, with an initial learning rate of 0.01.

\subsection{Periodic update of PV inflow model} \label{section:periodic_pv_update}
The PV inflow models use parameters fit to historical data. However, as the underlying system can change over time, depending on parameters we cannot directly measure (e.g. periodic cleaning of the PV array), we retrain the predictive models every time a new forecast is received (which occurs at midnight and noon each day). We retrain over data from the past 10 days. Separate models are fit for the forecasts at midnight and noon, leading to a total of four models.

\subsection{Comparison of PV output prediction methods}
We report the root mean square error (RMSE) of the one-day ahead prediction, updated every 12 hours as in Section \ref{section:periodic_pv_update}, relative to the true PV output. The model produces one prediction per forecast, so there are 21 different predictions. The reported RMSE is averaged across the 21 predictions. 

\begin{table}
\centering 
\begin{tabular}{| c | c | c |}
    \hline
     & PVUSA RMSE & ANN RMSE \\ 
    \hline
    Clear scenarios & $5.80$ &  $6.66$ \\ 
    \hline
	Cloudy scenarios & $8.38$ & $9.03$  \\ 
	\hline
    Combined scenarios & $7.26$ & $8.03$ \\
	\hline
\end{tabular}
\vspace*{2mm}
\caption{Obtained RMSE in \SI{}{\kilo \watt} for one-day-ahead prediction of PV output for clear and cloudy days using PVUSA and ANN prediction models, including a first-order error filter.}
\vspace*{-5mm}
\label{table:pv_errors_clear}
\end{table}

Table \ref{table:pv_errors_clear} shows the RMSE achieved by the two models for one-day-ahead prediction, considering clear, cloudy, and all (combined) scenarios.  We choose the PVUSA model for its improved predictive power and simplicity relative to the ANN model. Note that the ANN model is more flexible, and perhaps could result in improved performance with a different choice of predictors and model structure. As we are using weather forecasts as input for our models, the error between forecast and realized irradiance likely dominates the total model error.

\subsection{First-order filter on error} \label{section:first_order_filter}
To further decrease the prediction error of the PV production model, we implement a first-order filter. Suppose that at timestep $n$, we have predicted a PV production of $f_n$, realization $r_n$, and prediction error $e_n = f_n - r_n$. As the realization is not revealed until the end of the timestep, we instead compute $e_{n-1} = f_{n-1} - r_{n-1}$ from the previous timestep. We then estimate $e_n$ as $e^*_{n} = \alpha e_{n-1}$. Here, $\alpha$ is selected based on historical PV output prediction data. Our modified forecast is thus $f^*_n = f_n - \alpha e_{n-1}$. We use this correction for subsequent timesteps in the horizon as well, with $f^*_{n+i} = f_{n+i} - \alpha^{i+1} e_{n+i-1}$ for $i=0,\ldots,M-1$.

We choose $\alpha = 0.5$ by testing values on a grid between 0 and 1, and then comparing the RMSE of the resulting one-step-ahead (hour-ahead) prediction error achieved on historical data. Correcting the forecast using the first-order filter reduces the one-step-ahead RMSE from \SI{4.72}{\kilo \watt} to \SI{4.03}{\kilo \watt}.

\section{Building Electricity Demand Model} \label{section:demand_model}
To predict the building demand for the five day forecast horizon, we fit one ANN to the first day and another ANN to the subsequent four days.

We use cross-validation to choose the following ANN parameters for the day-ahead demand prediction:
\begin{enumerate*}[label=(\roman*)]
\item an input layer of width 60, corresponding to a scenario of 24 irradiance and 24 temperature forecasts, as well as the demand for the previous 12 hours,
\item two fully-connected layers of width 96 and 60 respectively, each with a ReLU activation function, and
\item an output layer of width 24, corresponding to the next 24 values of the demand.
\end{enumerate*}

To predict the demand corresponding to forecasts that are two to five days ahead, we use the following ANN structure:
\begin{enumerate*}[label=(\roman*)]
\item an input layer of width 48, corresponding to a scenario of 24 irradiance and 24 temperature forecasts,
\item a fully-connected layer of width 96, ReLU activation function and dropout layer with probability 45\%,
\item a fully-connected layer of width 72, ReLU activation function and dropout layer with probability 40\%, and
\item an output layer of width 24.
\end{enumerate*}

We normalize the irradiance and temperature forecasts, but do not normalize the demand, as an accurate upper limit is not known a priori. The hour-ahead prediction RMSE of this model improves from \SI{6.27}{\kilo \watt} to \SI{5.23 }{\kilo \watt} when using the first-order error correction of Section~\ref{section:first_order_filter} with $\alpha = 0.5$.

\section{Deterministic Peak Shaving Problem} \label{section:deterministic_problem}
We wish to minimize the electricity costs of a building connected to the electricity distribution grid, which is subject to a tariff consisting of a time-of-day energy charge $p^\mathrm{buy}$ and a monthly peak demand charge $p^\mathrm{peak}$. The building electricity demand $P^\mathrm{dem}$ must be met at each timestep. An attached PV array also supplies power $P^\mathrm{PV}$ to the building.

In a setting \emph{without} a BESS, the power $P^\mathrm{grid}$ purchased from the grid to meet the building demand at each timestep is simply the difference between the demand and PV production. However, here we consider a setting \emph{with} a BESS, leading to additional flexibility in choosing battery charging and discharging powers $P^\mathrm{C} \geq 0$ and $P^\mathrm{DC} \geq 0$. 

We consider a generic BESS, with limited charging and discharging powers $P^\mathrm{C,max}$ and $P^\mathrm{DC,max}$, and constant conversion efficiencies \cite{Park2017}. The change in stored energy $\Delta E$ in the BESS over a fixed time period is written as 
\begin{equation} \label{eq:storage_change_single}
\Delta E(P^\mathrm{C},P^\mathrm{DC}) = m^\mathrm{C} P^\mathrm{C} - 1/m^\mathrm{DC} P^\mathrm{DC},
\end{equation}
with $m^\mathrm{C}$ and $m^\mathrm{DC}$ the charging and discharging efficiencies.

The stored energy at the terminal stage is assessed a value based on the minimum purchase price:
\begin{equation} \label{eq:terminal_cost}
p^\mathrm{term} = \left( \min_t p_t^\mathrm{buy} \right) / m^\mathrm{C} 
\end{equation}

In Section \ref{section:optimal_deterministic}, we present a deterministic optimization problem where the PV output and electricity demand are known over the entire period of interest. This knowledge is clearly unrealistic in practice, but serves as a benchmark for subsequent algorithms that we develop. Section \ref{section:deterministic_mpc_formulation} then considers a deterministic MPC problem, where the problem data are known only over a limited horizon. This likewise unrealistic setting allows us to address how to account in the objective for multiple peak periods that may fall within the limited horizon. We can then proceed to the realistic, stochastic setting of Section \ref{section:stochastic_problem}, where information about the PV and demand comes solely from forecasts over a limited horizon.

\subsection{Optimal deterministic solution} \label{section:optimal_deterministic}
We initially formulate a deterministic optimization problem to operate the BESS, maximizing an economic objective while assuming that the problem data, including $p^\mathrm{buy}$, $p^\mathrm{peak}$, $P^\mathrm{dem}$, and $P^\mathrm{PV}$ are all given. We optimize over a horizon of length $N$ that is comprised of $Q$ peak periods.  The optimization problem is written as the following LP:
\begin{subequations} \label{eq:deterministic_problem}
\begin{align} \label{eq:deterministic_objective}
  \min \ & \sum_{t=0}^{N-1} p_t^\mathrm{buy} P_t^\mathrm{grid} + p^\mathrm{peak} \sum_{q=1}^Q s_q - p^\mathrm{term} E_N
\\ \label{eq:deterministic_power_balance}
\text{s.t.\quad} & P_t^\mathrm{grid} \geq P_t^\mathrm{dem} - P_t^\mathrm{PV} + P_t^\mathrm{C} - P_t^\mathrm{DC}
\\ \label{eq:deterministic_storage_dynamics}
& E_{t+1} = E_{t} + \Delta E (P_t^\mathrm{C}, P_t^\mathrm{DC}) 
\\ \label{eq:deterministic_volume_limits}
& 0 \leq E_t \leq E^\mathrm{max}
\\
& P_t^\mathrm{grid} \geq 0 
\\
& 0 \leq P_t^\mathrm{C} \leq P^\mathrm{C,max}  
\\ \label{eq:deterministic_dc_ub}
& 0\leq P_t^\mathrm{DC} \leq P^\mathrm{DC,max}  
\\ \label{eq:peak_calculation}
& s_q=\max \{P_k^\mathrm{grid} \ | \ k \in \text{peak period} \ q \} 
\\
& E_0 \ \text{given,}
\end{align}
\end{subequations}
where \eqref{eq:deterministic_power_balance}-\eqref{eq:deterministic_dc_ub} hold for all $t = 0,\ldots,N-1$ and \eqref{eq:peak_calculation} holds for all $q=1,\ldots,Q$.

In the objective \eqref{eq:deterministic_objective}, the storage operator incurs energy and peak demand charges, with a per-unit value $p^\mathrm{term}$ from \eqref{eq:terminal_cost} ascribed to the stored energy $E_N$ at the terminal stage. 

The power balance between the grid, battery, demand, and PV inflow is specified in \eqref{eq:deterministic_power_balance}. Note the  inequality, which accounts for timesteps where the PV production is higher than the demand. In such cases, we assume that the PV inflow can be curtailed if needed (e.g., when the storage is full).

The dynamics of the stored energy $E_t$ are given in \eqref{eq:deterministic_storage_dynamics}, with $\Delta E$, the change in stored energy, a function of the charging and discharging powers as in \eqref{eq:storage_change_single}. The stored energy is nonnegative and bounded above by $E^\mathrm{max}$ in \eqref{eq:deterministic_volume_limits}.

Finally, the peak usage $s_q$ in each peak period is calculated as in \eqref{eq:peak_calculation}. This pointwise maximum across all timesteps in a peak period is implemented using an additional epigraph variable, resulting in a linear constraint.

\subsection{Deterministic MPC setting} \label{section:deterministic_mpc_formulation}
The assumption that problem data is known for the entire horizon is unrealistic, due to the inherent uncertainty in the PV inflow and building demand. As a first step towards our goal of making real-time decisions to solve the true underlying stochastic problem, we consider the deterministic problem over a finite MPC horizon of length $M$. For simplicity, we assume that $M$ is shorter than the peak period length $\ell^\mathrm{peak}$, meaning we consider at most two peak periods, with peak usage $s_u$ and $s_{u+1}$, in a given horizon. Here, $u \in \{1,\ldots,Q\}$ is the peak period index. If $M > \ell^\mathrm{peak}$, the following can be extended in a straightforward manner to incorporate peak periods $s_{u+2}$, $s_{u+3}$, etc. 

At each timestep $t = 0,\ldots,N-1$, we receive the latest problem data and forecasts, and then solve the LP
\begin{subequations} \label{eq:deterministic_mpc_problem}
\begin{align} \label{eq:mpc_objective}
  \min \ & \sum_{k=t}^{t+M-1} \! p_k^\mathrm{buy} P_k^\mathrm{grid} + \frac{M}{\ell^\mathrm{peak}} f^\mathrm{peak}(s_u,s_{u+1}) - p^\mathrm{term} E_{t+M}
\\ \label{eq:mpc_power_balance}
\text{s.t.\quad} & P_k^\mathrm{grid} \geq P_k^\mathrm{dem} - P_k^\mathrm{PV} + P_k^\mathrm{C} - P_k^\mathrm{DC}
\\ \label{eq:mpc_energy_integration}
& E_{k+1} = E_{k} + \Delta E (P_k^\mathrm{C}, P_k^\mathrm{DC}) 
\\
& 0 \leq E_k \leq E^\mathrm{max}
\\
& P_k^\mathrm{grid} \geq 0 
\\
& 0 \leq P_k^\mathrm{C} \leq P^\mathrm{C,max}  
\\ \label{eq:deterministic_mpc_dc_ub}
& 0\leq P_k^\mathrm{DC} \leq P^\mathrm{DC,max}  
\\ \label{eq:deterministic_mpc_peak_calculation}
& s_q \geq P_k^\mathrm{grid} \ \text{if} \ k \in \text{peak period} \ q 
\\ \label{eq:mpc_initial_peak_assignment}
& s_u \geq s_u^\mathrm{init}
\\ \label{eq:deterministic_mpc_initial_conds}
& E_t, s_u^\mathrm{init} \ \text{given,}
\end{align}
\end{subequations}
where \eqref{eq:mpc_power_balance}-\eqref{eq:deterministic_mpc_peak_calculation} hold for all $k = t,\ldots,t+M-1$ and \eqref{eq:deterministic_mpc_peak_calculation} additionally holds for $q = u, u+1$. Here, $u$ is the index of the peak period to which timestep $t$ belongs. After solving the LP at each timestep, we apply the computed optimal actions $P_t^\mathrm{grid}$, $P_t^\mathrm{C}$, and $P_t^\mathrm{DC}$ for the current timestep in an MPC fashion.

At timestep $t$, the relevant problem state consists of the current storage level $E_t$ and maximum grid power $s_u^\mathrm{init}$ seen thus far in the current peak period $u$. 

The main difference between problems \eqref{eq:deterministic_problem} and \eqref{eq:deterministic_mpc_problem} is that in \eqref{eq:deterministic_mpc_problem} we solve over a shorter horizon. While \eqref{eq:deterministic_problem} considers all timesteps $0,\ldots,N$, in \eqref{eq:deterministic_mpc_problem} we only consider the maximum grid power up to timestep $t+M-1$. Depending on the MPC horizon length and particular timestep considered, the horizon can either overlap with multiple peak periods (with peak grid usage $s_u$ and $s_{u+1}$), or else be contained within the current peak period (with peak usage $s_u$). 

The weighting factor $M / \ell^\mathrm{peak}$ in the objective \eqref{eq:mpc_objective} serves to correct the relative weighting between the peak cost $f^\mathrm{peak}(s_u,s_{u+1})$ and the energy cost $\sum_{k=t}^{t+M-1} p_k^\mathrm{buy} P_k^\mathrm{grid}$, which would otherwise be skewed for horizon lengths different than a whole peak period. 

We penalize the peak cost in the objective function via $f^\mathrm{peak}(s_u,s_{u+1})$. Note that in the MPC setting, although the decision taken each timestep only affects the current peak period, a prudent optimization strategy must still account for the peak cost incurred in the future peak period. Here, we consider three potential formulations that account for this. 
\begin{enumerate}[label=(\alph*)]
\item $f^\mathrm{peak}(s_u,s_{u+1}) = p^\mathrm{peak} \left( s_u+ s_{u+1} \right)$, with the peak grid power in each peak period penalized equally.  
\item $f^\mathrm{peak}(s_u,s_{u+1}) = p^\mathrm{peak} \left(\beta s_u+(1-\beta)s_{u+1}\right)$, where $\beta \in [0,1]$ is the fraction of the timesteps of the MPC horizon of length $M$ which fall in the current peak period, and, consequently, $1-\beta$ the fraction that fall in the subsequent peak period. In this case, each peak is penalized depending on the point in the horizon where the MPC problem is solved.
\item $f^\mathrm{peak}(s_u,s_{u+1}) = p^\mathrm{peak} \left(s_u+(1-\beta)s_{u+1}\right)$, where $\beta$ is as above. In this case, the peak in the current period is fully penalized, and the peak in the subsequent period is partially penalized.
\end{enumerate}

\section{Stochastic Peak Shaving Problem} \label{section:stochastic_problem}
We now propose a stochastic formulation of the peak shaving problem, where information about $P^\mathrm{dem}$ and $P^\mathrm{PV}$ comes solely from MeteoSwiss forecast scenarios of irradiance and temperature. Using the models developed in Sections \ref{section:pv_inflow_model} and \ref{section:demand_model}, we generate forecasts of $P_k^{\mathrm{dem}, \, j}$ and $P_k^{\mathrm{PV}, \, j}$ corresponding to each forecast scenario $j$ and timestep $k$. To simplify subsequent discussion, we consolidate the two forecasts into a forecast of net demand 
\begin{equation} \label{eq:net_demand_eq}
P_k^{\mathrm{dem,net}, \, j} = P_k^{\mathrm{dem}, \, j} -P_k^{\mathrm{PV}, \, j}.
\end{equation}

At timestep $t$, the stochastic problem is formulated as:
\begin{subequations} \label{eq:stochastic_problem}
\begin{align} \nonumber
  \min  \sum_{j=1}^{21} &  \sum_{k=t}^{t+M-1} p_k^\mathrm{buy} P_k^{\mathrm{grid}, \, j}  +  \frac{M}{\ell^\mathrm{peak}} f^\mathrm{peak}(s^j_u,s^j_{u+1})
\\  \label{eq:stochastic_objective}
& -  p^\mathrm{term} E_{t+M}^j
\\ \label{eq:stochastic_mpc_power_balance_simplified}
\text{s.t.\quad} & P_k^{\mathrm{grid}, \, j} \geq P_k^{\mathrm{dem,net}, \, j} + P_k^{\mathrm{C}, \, j} - P_k^{\mathrm{DC}, \, j}
\\ \label{eq:stochastic_mpc_energy_integration}
& E_{k+1}^j = E_k^j + \Delta E (P_k^{\mathrm{C}, \, j}, P_k^{\mathrm{DC}, \, j}) 
\\
& 0 \leq E_k^j \leq E^\mathrm{max}
\\
& P_k^{\mathrm{grid}, \, j} \geq 0 
\\
& 0 \leq P_k^{\mathrm{C}, \, j} \leq P^\mathrm{C,max}  
\\ \label{eq:stochastic_mpc_dc_ub}
& 0\leq P_k^{\mathrm{DC}, \, j} \leq P^\mathrm{DC,max}  
\\ \label{eq:stochastic_mpc_peak_calculation}
& s_q^j \geq P_k^{\mathrm{grid}, \, j} \ \text{if} \ k \in \text{peak period} \ q 
\\ \label{eq:stochastic_mpc_initial_peak_assignment}
& s_u^j \geq s_u^\mathrm{init}
\\ \label{eq:stochastic_mpc_initial_conds}
& E_t, s_u^\mathrm{init} \ \text{given,}
\end{align}
\end{subequations}
where \eqref{eq:stochastic_mpc_power_balance_simplified}-\eqref{eq:stochastic_mpc_peak_calculation} hold for all timesteps $k = t,\ldots,t+M-1$ and scenarios $j=1,\ldots,21$. Additionally, \eqref{eq:stochastic_mpc_peak_calculation} holds for peak periods $q = u,u+1$, and \eqref{eq:stochastic_mpc_initial_peak_assignment} holds for scenarios $j=1,\ldots,21$, since the initial conditions $E_t$ and $s_u^\mathrm{init}$ are shared across all scenarios.

The stochastic optimal control problem \eqref{eq:stochastic_problem} minimizes a cost which is summed over all scenarios $j=1,\ldots,21$. Our goal in the MPC setting is to choose the decision variables to apply for timestep $t$. While these must be consistent between scenarios at timestep $t$, scenario-dependent decisions are possible for subsequent timesteps (where $k \geq t+1$), since they will not be applied in the MPC setting.  

We now present several methods which produce scenario-independent decision policies for $P_t^\mathrm{grid}$. As discussed in Section \ref{section:infeasibility_handling}, when $P_t^\mathrm{dem,net}$ is known and $P_t^\mathrm{grid}$ chosen, the optimal choice of $P_t^\mathrm{C}$ and $P_t^\mathrm{DC}$ follows from \eqref{eq:stochastic_mpc_power_balance_simplified}.

\subsection{MPC with policy for initial timestep} \label{section:causal_stochastic_mpc}
We assume that the realization of the uncertainty $P_t^\mathrm{dem,net}$ is not available at the time the optimization problem is \emph{solved}, but is available at the time the control inputs are \emph{applied}. Therefore, the applied decision $P_t^\mathrm{grid}$ can depend on the unknown realization of $P_t^\mathrm{dem,net}$ via a policy. When the realization becomes available, the policy can be used to compute the $P_t^\mathrm{grid}$ to be applied.

The policy we find for $P_t^\mathrm{grid}$ must be the same for all scenarios, so that it results in a unique choice of $P_t^\mathrm{grid}$ for a particular realization of $P_t^\mathrm{dem,net}$. The policy must be feasible for all scenarios $P_t^{\mathrm{dem,net}, \, j}$, but need not necessarily be feasible for all possible $P_t^\mathrm{dem,net}$. 

For the time being, we allow the choice of $P_k^{\mathrm{grid}, \, j}$ (and other decision variables) for timesteps $k = t+1,\ldots,t+M-1$ to vary freely across scenarios. Note that this lack of coupling between scenarios is overly optimistic due to the dependence of variables on individual scenarios. This policy parametrization is revisited in Section \ref{section:causal_timesteps}.  

We now consider two policies for the applied decision $P_t^\mathrm{grid}$, as a function of $P_t^\mathrm{dem,net}$. 

\subsubsection{Decision without knowledge of uncertainty}
A simple policy is to assume that a single $P_t^\mathrm{grid}$ will be applied at time $t$, regardless of the realization of the uncertainty $P_t^\mathrm{dem,net}$. To find such a policy, we solve the stochastic optimization problem \eqref{eq:stochastic_problem} with the additional constraint
\begin{equation} \label{eq:first_stage_same_constraint}
	P^{\mathrm{grid}, \, j}_t = P^\mathrm{grid, 1}_t, \ j = 2,\ldots,21
\end{equation}
to ensure the grid power at time $t$ is the same for all scenarios. Since $P_t^\mathrm{grid}$ is the same for all scenarios, but $P_t^{\mathrm{dem,net}, \, j}$ varies across scenarios, this results in scenario-dependent $P^{\mathrm{C}, \, j}_t$, $P^{\mathrm{DC}, \, j}_t$, $s_u^j$, and $E^j_t$.

\subsubsection{Saturated affine decision policy} \label{section:affine_decision_policy}
A policy for $P_t^\mathrm{grid}$ that depends on $P_t^\mathrm{dem,net}$ can provide additional flexibility. From \cite{Lygeros2011}, we consider an affine decision policy of the form 
\begin{equation} \label{eq:causal_policy_structure}
P_t^\mathrm{grid} = a_t P_t^\mathrm{dem,net} + b_t,
\end{equation}
where $a_t \in \mathbb{R}$ and $b_t \in \mathbb{R}$ are optimization variables. 

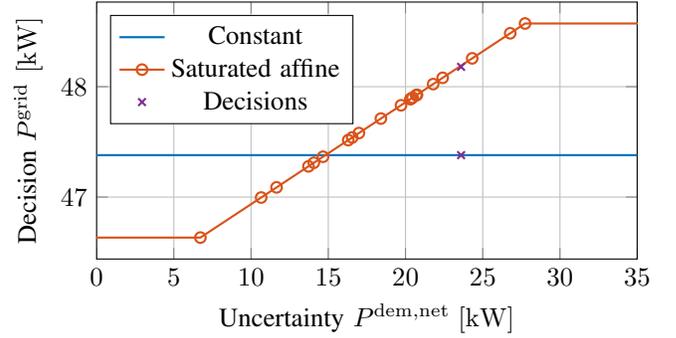
\begin{figure}
    \centering
\begin{tikzpicture}
 \begin{groupplot}[
     group style = {group size = 1 by 1},
     height = 5 cm,
     xmin = 0, 
     xmax = 35,
     width = 0.99\linewidth,
     grid=both
    ]
    \nextgroupplot[
     grid=both,
     xlabel={Uncertainty $P^\mathrm{dem,net} \ [\SI{}{\kilo \watt}]$},
     ylabel={Decision $P^\mathrm{grid} \ [\SI{}{\kilo \watt}]$},
     legend style={at={(0.25,0.95)}, anchor=north}
    ]
    \addplot[matlabblue, thick] coordinates {
        (0,47.3798)
	    (35,47.3798)
		};
    \addplot[matlabred,thick, mark=o] coordinates {
        (-2,46.6326)
        (6.7201,46.6326)
        (10.6552,46.9957)
        (11.6529,47.0878)
        (13.7178,47.2784)
        (14.066,47.3106)
        (14.6587,47.3653)
        (16.291,47.5159)
        (16.5506,47.5399)
        (16.984,47.5799)
        (18.4065,47.7112)
        (19.7106,47.8315)
        (20.3174,47.8875)
        (20.3483,47.8904)
        (20.4614,47.9008)
        (20.6914,47.9221)
        (20.7443,47.9269)
        (21.7905,48.0235)
        (22.406,48.0803)
        (24.3183,48.2568)
        (26.7797,48.484)
        (27.7405,48.5727)
        (37,48.5727)  
		};
	\addplot[matlabpurple,thick, mark=x, only marks] coordinates {
        (23.6,47.3798)
        (23.6,48.18) 
		};
    
		\legend{Constant, Saturated affine, Decisions}
  \end{groupplot}
\end{tikzpicture}
    \caption{Comparison of constant and saturated affine decision policies for a given timestep. Affine decision policy is plotted with circles denoting the forecast $P^\mathrm{dem,net}$ for the 21 scenarios. The actual realization of the uncertainty is $\SI{23.6}{\kilo \watt}$, leading to a decision of $\SI{47.4}{\kilo \watt}$ for the constant decision policy, and $\SI{48.2}{\kilo \watt}$ for the affine decision policy.}
    \label{fig:linear_decision_policy}
\end{figure}

We solve \eqref{eq:stochastic_problem} with the additional constraint
\begin{equation} \label{eq:decision_policy_structure_scen}
P_t^{\mathrm{grid}, \, j} = a_t P_t^{\mathrm{dem,net}, \, j} + b_t, \ j = 1,\ldots,21
\end{equation}
and additional decision variables $a_t$ and $b_t$ to derive the same policy for all scenarios. When the true $P_t^\mathrm{dem,net}$ is revealed, the policy is evaluated using \eqref{eq:causal_policy_structure} to determine $P_t^\mathrm{grid}$. Realizations outside the forecast range use nearest neighbor interpolation, thereby saturating the decision. Fig.~\ref{fig:linear_decision_policy} compares example constant and saturated affine decision policies.

% above data is for first timestep, Mar. 1 - Apr. 30

\subsection{Modification of scenarios in first timestep}
The forecast scenarios $P_t^{\mathrm{dem,net}, \, j}$ are often tightly clustered.  
To increase the likelihood that the realization of $P_t^\mathrm{dem,net}$ falls within the forecast scenario range, we add a Gaussian random variable $\omega_t \sim \mathcal{N}(0,\epsilon_t^2)$ to the scenarios for the initial timestep. That is, we modify \eqref{eq:net_demand_eq} for the initial timestep $t$ as 
\begin{equation} \label{eq:net_demand_eq_mod}
P_t^{\mathrm{dem,net}, \, j} = P_t^{\mathrm{dem}, \, j} - P_t^{\mathrm{PV}, \, j} + \omega_t
\end{equation}
The standard deviation $\epsilon_t$ is set as 
\begin{equation} \label{eq:first_timestep_noise}
\epsilon_t = \max(P^\mathrm{dem,net}_\mathrm{err} - \range(P^\mathrm{dem,net}_\mathrm{t}) ,0),
\end{equation}
where $P^\mathrm{dem,net}_\mathrm{err}$ is the historical RMSE in predicting $P^\mathrm{dem,net}$, and $\range(P^\mathrm{dem,net}_\mathrm{t}) = \max_j P_t^{\mathrm{dem,net}, \, j} - \min_j P_t^{\mathrm{dem,net}, \, j}$ is the range of the scenarios in the first timestep. 
%The effects of this modification are discussed in Section \ref{section:stochastic_mpc_results}.

Scenarios are not modified for timesteps $t+1,\ldots,t+M-1$, since the decisions made at timestep $t$ for subsequent timesteps are not applied in the MPC setting.

\subsection{Relative weighting within peak}
Forecast accuracy generally decreases the farther in the future a forecast is made. Inspired by this, we propose a modified peak penalty $f^\mathrm{peak*}(s_u,s_{u+1})$ that includes a term that considers the peak over the initial $M_1$ timesteps in addition to the peak over the entire horizon. We weight the first term by a parameter $\theta \in [0,1]$, so that
\begin{equation} \label{eq:stochastic_mpc_subpeak_weighting}
	f^\mathrm{peak*} \! (s_u,s_{u+1}) \! = \! \theta f^\mathrm{peak}_{t \leq M_1}(s_u,s_{u+1}) + (1-\theta) f^\mathrm{peak}(s_u,s_{u+1}).
\end{equation}
For example, choosing $M_1 = 24$ means that we add a term that penalizes the peak over the next 24 hours (in addition to the peak over the entire MPC horizon of length $M$).

\subsection{Policy evaluation, infeasibility} \label{section:infeasibility_handling}
The above policies determine $P_t^\mathrm{grid}$ as a function of the realization of $P^\mathrm{dem,net}_\mathrm{t}$. Given $P^\mathrm{dem,net}_\mathrm{t}$ and $P_t^\mathrm{grid}$, we now wish to find the resulting optimal choices for $P^\mathrm{C}_t$ and $P^\mathrm{DC}_t$.

Only one of the charging and discharging powers can be nonzero, due to the positive electricity prices and lossy conversion efficiencies used here. We initially assign the net power $\lvert P_t^\mathrm{grid} - P^\mathrm{dem,net}_\mathrm{t} \rvert$ to either $P^\mathrm{C}_t$ or $P^\mathrm{DC}_t$, depending on the sign of $P_t^\mathrm{grid} - P^\mathrm{dem,net}_\mathrm{t}$.

Unfortunately, this may result in an infeasible solution, if, for example, the realization of $P^\mathrm{dem,net}_\mathrm{t}$ lies outside of the range of the forecast uncertainty, and the charging or discharging power limits are violated. If problem constraints are violated due to too much power being present, we can curtail the PV production present in $P^\mathrm{dem,net}_\mathrm{t}$ as necessary. If too little power is present, then $P_t^\mathrm{grid}$ can be increased. The grid and PV curtailment thus provide slack. 

\subsection{Treatment of subsequent timesteps in horizon} \label{section:causal_timesteps}
In the previous sections, the decisions made for timesteps $t+1,\ldots,t+M-1$ can be chosen independently for different scenarios. This can lead to optimistic behavior, since current decisions are allowed to depend on the forecasts of timesteps in the future. To combat this, we look for a policy of the form
\begin{equation} \label{eq:causal_policy_structure2}
P^\mathrm{grid} = A P^\mathrm{dem,net} + B
\end{equation}
where $P^\mathrm{grid} = [P_t^\mathrm{grid},\ldots,P_{t+M-1}^\mathrm{grid}]^\top$ and $P^\mathrm{dem,net} = [P_t^\mathrm{dem,net}, \ldots, P_{t+M-1}^\mathrm{dem,net}]^\top$. Here, $B$ is a vector of length $M$, and $A$ is a lower triangular matrix of dimension $M \times M$. 

Following \cite{Lygeros2011}, the requirement that $A$ is lower triangular ensures a \emph{causal} policy, since the decision at time $k$ depends only on information from timesteps $t,\ldots,k$. To reduce computational load, we impose a lower banded structure on $A$, where, for each row $i$, we impose the additional constraints
\begin{equation} \label{eq:banded_causal_policy}
A_{ij} = 0 \ \textrm{if } j < i - M_2
\end{equation}
for some positive integer $M_2$. In this case, the number of decision variables grows linearly in $M$.

\section{Simulation results and analysis} \label{section:results}
We wish to operate a generic BESS attached to a building, with the building demand and PV inflow as described in Section \ref{section:models_and_data}. The \SI{2500}{\kilo \watt \hour} BESS has maximum charging and discharging powers of \SI{100}{\kilo \watt}. Conversion efficiencies in \eqref{eq:storage_change_single} are set to $m^\mathrm{C} = m^\mathrm{DC} = 90\%$.

For operating costs, we consider the \textit{400LS} tariff for commercial customers connected to the distribution grid in the canton of Zurich for 2018 \cite{EKZ_price_legacy}. The tariff consists of the time-of-use rate $p_t^\mathrm{buy}$ of 13.98 cents during the day and 9.33 cents during the night, as well as a monthly demand charge of $p^\mathrm{peak} = \SI{3.05}{CHF \per \kW}$ of peak usage. The terminal storage volume $E_N$ is assessed a value based on \eqref{eq:terminal_cost}, with the minimum price taken over the current optimization horizon.

We present simulation results for the various optimization problems formulated in Sections \ref{section:deterministic_problem} and \ref{section:stochastic_problem}. We use the problem data of Section \ref{section:models_and_data}, from March 1 until April 30, 2018 (skipping March 3 and 18 due to missing data). Timesteps have a length of one hour to match the forecast granularity.

\subsection{Optimal deterministic results}
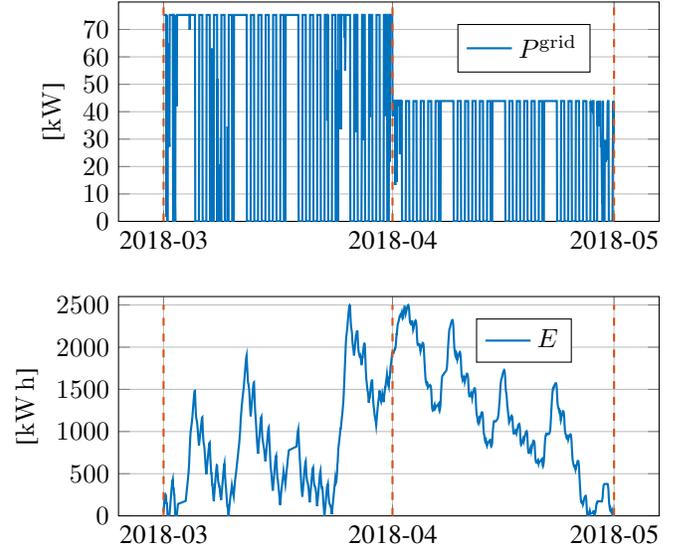
\begin{figure}
    \centering
\begin{tikzpicture}
 \begin{groupplot}[
     group style = {group size = 1 by 2},
     height = 4.5 cm,
     width = 0.99\linewidth,
     grid=both
    ]
    \nextgroupplot[ylabel={[$\SI{}{\kilo \watt}$]},
                  ymin = 0,
                  ymax = 80,
                  ytick={0, 10, 20, 30, 40, 50, 60, 70},
                  /pgf/number format/.cd,
                  use comma,
                  1000 sep={},
                  legend style={at={(0.75,0.9)},
		anchor=north},
		date coordinates in=x, date ZERO=2018-01-01, xticklabel=\year-\month,xtick = {2018-03-01, 2018-04-01, 2018-05-01}]
      \addplot[matlabblue, no marks, thick] table [x = timestamp, y = p_grid_buy_opt, col sep=comma] {optimal_control_data_mar_apr.csv};
      \addplot [matlabred, dashed, thick] coordinates {
	    (2018-03-01,0)
	    (2018-03-01,80)};
      \addplot [matlabred, dashed, thick] coordinates {
	    (2018-04-01,0)
	    (2018-04-01,80)};
	  \addplot [matlabred, dashed, thick] coordinates {
	    (2018-05-01,0)
	    (2018-05-01,80)};
      \legend{$P^\mathrm{grid}$}
  \nextgroupplot[ylabel={[$\SI{}{\kilo \watt \hour}$]},
		          ymin = 0,
		          ymax = 2600,
		          ytick={0, 500, 1000, 1500, 2000, 2500},
                  /pgf/number format/.cd,
                  use comma,
                  1000 sep={},
                  legend style={at={(0.75,0.9)},
		anchor=north},
		date coordinates in=x, date ZERO=2018-01-01, xticklabel=\year-\month,xtick = {2018-03-01, 2018-04-01, 2018-05-01}]
      \addplot[matlabblue, no marks, thick] table [x = timestamp, y = v_stored_opt, col sep=comma] {optimal_control_data_mar_apr.csv};
      \addplot [matlabred, dashed, thick] coordinates {
	    (2018-03-01,0)
	    (2018-03-01,2500)};
      \addplot [matlabred, dashed, thick] coordinates {
	    (2018-04-01,0)
	    (2018-04-01,2500)};
	  \addplot [matlabred, dashed, thick] coordinates {
	    (2018-05-01,0)
	    (2018-05-01,2500)};
      \legend{$E$}
  \end{groupplot}
\end{tikzpicture}
    \caption{Optimal trajectory of deterministic problem with power purchased from grid $P^\mathrm{grid}$ and stored energy $E$. Peak period boundaries are in red.}
    \label{fig:optimal_deterministic_trajectory}
\end{figure}

As a benchmark, we solve \eqref{eq:deterministic_problem}, which has perfect knowledge of the problem data over the entire problem horizon. The resulting behavior is presented in Fig.~\ref{fig:optimal_deterministic_trajectory}. Due to the perfect foresight, the peak value of purchased power in each peak period is achieved near the beginning of the period. The optimal objective for the deterministic peak shaving problem is 4848.98 CHF, of which 363.51 CHF is from the peak penalty. 

\subsection{Deterministic MPC results} \label{section:deterministic_mpc_results}
We next determine optimal parameters for the deterministic MPC peak shaving problem \eqref{eq:deterministic_mpc_problem}. We first consider the three different peak period weighting strategies $f^\mathrm{peak}(s_u,s_{u+1})$ that were presented in Section \ref{section:deterministic_mpc_formulation}. As a reminder, these were
\begin{itemize}
    \item 100\% fixed weight on first period, proportional weight on second period, denoted here as ``F+P'' 
    \item 100\% fixed weight on both periods, denoted as ``F+F'' 
    \item Each period weighted proportionally, denoted as ``P+P'' 
\end{itemize}

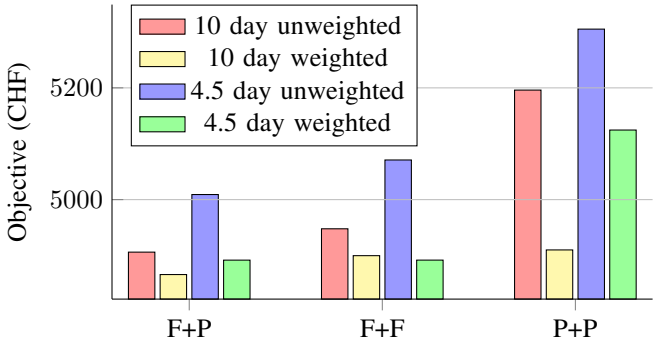
\begin{figure}
    \centering
\begin{tikzpicture}
\begin{axis}[/pgf/number format/.cd,
                  use comma,
                  1000 sep={},
legend style={at={(0.3,1.0)},
		anchor=north},
every axis plot post/.style={/pgf/number format/fixed},
ybar=2pt,
bar width=10pt,
enlarge x limits=0.2,
visualization depends on=rawy\as\rawy, % Save the unclipped values
after end axis/.code={ % Draw line indicating break
\draw [ultra thick, white] (rel axis cs:0,1.05) -- (rel axis cs:1,1.05);
    },
clip=false,
area legend,
height = 5.5 cm,
width = 0.99\linewidth,
axis on top,
ymajorgrids=true,
xtick=data,
ylabel=Objective (CHF),
symbolic x coords={F+P, F+F, P+P},
axis lines*=left
]
\addplot[fill=red!40] coordinates {(F+P,4906.0310) (F+F,4947.7973) (P+P,5196.0090) };
\addplot[fill=yellow!40] coordinates {(F+P,4865.9002) (F+F,4899.7084) (P+P,4909.9655)};
\addplot[fill=blue!40] coordinates {(F+P,5009.0984) (F+F,5070.9381) (P+P,5305.1162)};
\addplot[fill=green!40] coordinates {(F+P,4891.7174) (F+F,4891.7514) (P+P,5124.5239)};
\legend{10 day unweighted, 10 day weighted, 4.5 day unweighted, 4.5 day weighted};
\end{axis}
\end{tikzpicture}

\caption{Comparison of peak weighting strategies for deterministic MPC, including relative weighting of multiple horizon peaks, as well as relative weighting between peak cost and energy cost.}
\label{fig:peak_weighting_comparison}
\end{figure}

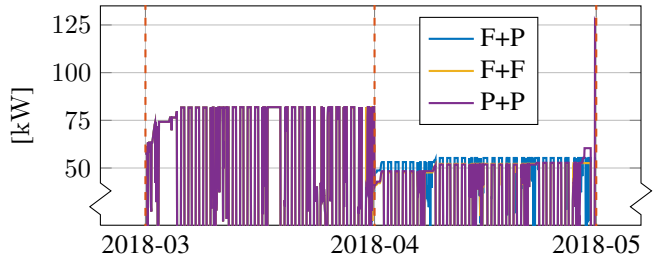
\begin{figure}
    \centering
\begin{tikzpicture}
 \begin{groupplot}[
     group style = {group size = 1 by 1},
     height = 4.5 cm,
     ymin = 20,
     ymax = 135,
     ytick={50, 75, 100, 125},
     width = 0.99\linewidth,
     grid=major
    ]  
    \nextgroupplot[ylabel={[$\SI{}{\kilo \watt}$]},
                  /pgf/number format/.cd,
                  use comma,
                  1000 sep={},
                  axis y discontinuity=crunch,
                  legend style={at={(0.70,0.95)},
		anchor=north},
		date coordinates in=x, date ZERO=2018-01-01, xticklabel=\year-\month,xtick = {2018-03-01, 2018-04-01, 2018-05-01}]
      \addplot[matlabblue, no marks, thick] table [x = timestamp, y = p_grid_buy_det_4d5_f+p, col sep=comma] {optimal_control_data_mar_apr.csv};
      \addplot[matlabyellow, no marks, thick] table [x = timestamp, y = p_grid_buy_det_4d5_f+f, col sep=comma] {optimal_control_data_mar_apr.csv};
      \addplot[matlabpurple, no marks, thick] table [x = timestamp, y = p_grid_buy_det_4d5_p+p, col sep=comma] {optimal_control_data_mar_apr.csv};
      \addplot [matlabred, dashed, thick] coordinates {
	    (2018-03-01,0)
	    (2018-03-01,2500)};
      \addplot [matlabred, dashed, thick] coordinates {
	    (2018-04-01,0)
	    (2018-04-01,2500)};
	  \addplot [matlabred, dashed, thick] coordinates {
	    (2018-05-01,0)
	    (2018-05-01,2500)};
      \legend{F+P, F+F, P+P}

  \end{groupplot}
\end{tikzpicture}
    \caption{Grid inflow power for deterministic MPC problem with 4.5 day horizon, considering three different peak relative weightings. Objective includes peak cost weighting factor as in \eqref{eq:mpc_objective}. Peak period boundaries are in red.}
    \label{fig:deterministic_mpc_weight_comparison}
\end{figure}

We solve the deterministic MPC problem for horizons of length 10 and 4.5 days, as well as different peak period weighting strategies and use of the weighting factor $M/ \ell^\mathrm{peak}$ (``weighted'' when present, ``unweighted'' when absent). The results are presented in Fig.~\ref{fig:peak_weighting_comparison}, with trajectories for the weighted 4.5 day horizon plotted in Fig.~\ref{fig:deterministic_mpc_weight_comparison}.

We note two points. First, the inclusion of the weighting factor $M/ \ell^\mathrm{peak}$ improves the objective, especially when considering the shorter 4.5 day MPC horizon. The peak cost is accrued over a month period, so for a proper tradeoff, the peak cost should be scaled to the fraction of the month it occupies.

Second, the ``F+P'' peak weighting strategy performs best in simulation among the various strategies proposed above. The ``F+P'' and ``F+F'' strategies differ in the degree of discontinuity introduced when the second peak appears in the objective. The ``F+F'' strategy encounters the new term in the objective in its entirety as soon as the end of the MPC horizon reaches the new peak period. For the 10 day horizon, this results in a slightly higher cost. The ``F+P'' strategy encounters the discontinuity from the new peak more gradually. 

The ``P+P'' strategy suffers from another flaw. As the MPC horizon nears the end of the current peak, less weight is placed on the current peak cost. This causes the peak value to rise at the end of the period, leading to a 4.8\% higher cost than when using the ``F+P'' strategy for a 4.5 day horizon. This effect depends on the problem data, as such a rise does not occur at the end of the first month.

\begin{figure}
    \centering
\begin{tikzpicture}
 \begin{axis}[
     xmin = 0, 
     xmax = 10.5,
     height = 4.5 cm,
     width = 0.99\linewidth,
     grid=major,
     thick,
     /pgf/number format/.cd,
     use comma,
     1000 sep={},
     xlabel={Horizon length $M$ [days]},
     ylabel={Objective [CHF]},
     legend style={at={(0.7,0.9)},
		anchor=north}
    ]
    \addplot[matlabblue,thick,mark=o] coordinates {
        (0.5,5246.7151)
        (1,5084.1500)
        (2,5002.3532)
        (3,5018.8607)
		(4,4895.3781)
		(4.5,4891.7174)
		(5,4880.3703)
		(10,4865.9002)
		};
    \addplot[black,densely dashed,thick] coordinates {
        (0,4848.98)
		(12,4848.98)
		};
		
      \legend{Deterministic MPC, Optimum}
  \end{axis}
\end{tikzpicture}
    \caption{Objective of deterministic MPC problem \eqref{eq:deterministic_mpc_problem} with F+P peak weighting strategy, as a function of horizon length $M$. Optimal solution solves \eqref{eq:deterministic_problem} for entire time horizon from March 1 - April 30, 2018.}
    \label{fig:deterministic_mpc_vs_horizon_length}
\end{figure}
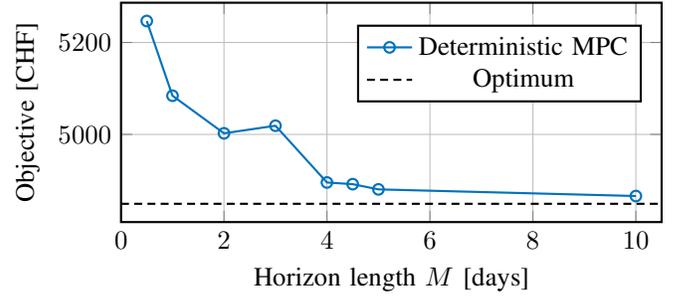

Finally, we wish to choose a horizon length $M$ for the MPC problem. As the horizon length increases, more information is available, but computational costs increase and forecast accuracy decreases. Using the ``F+P'' peak weighting strategy that was found to be beneficial above, we vary the horizon length in \eqref{eq:deterministic_mpc_problem}, and plot the achieved objective in Fig.~\ref{fig:deterministic_mpc_vs_horizon_length}. Since the provided forecasts are limited to 5 days, and are updated every 12 hours, we choose a horizon of 4.5 days so that the MPC horizon always remains within the given forecast.

\subsection{Stochastic MPC results} \label{section:stochastic_mpc_results}
We consider a stochastic MPC setting which uses the problem parameters and predictive models found to perform best in the deterministic MPC setting. These include
\begin{itemize}
	\item PV forecasting using a PVUSA model with parameters fit over data from the past 10 days and updated online, as well as a first-order filter on the error (with $\alpha = 0.5$)
	\item Demand forecasting using a multilayer ANN model fit over data from the past 10 days and updated online, as well as a first-order filter on the error (with $\alpha = 0.5$)
	\item 4.5 day MPC horizon ($M = 108$) with ``F+P'' peak period penalization strategy and objective containing the weighted peak cost. 
\end{itemize} 

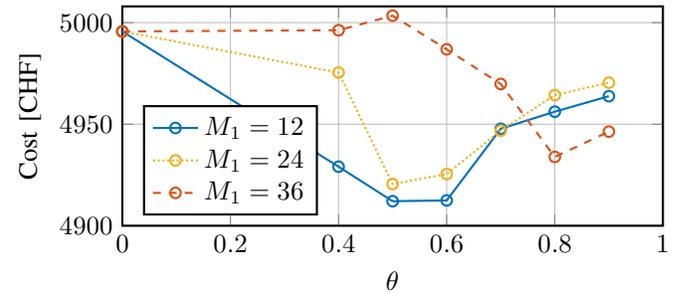
\begin{figure}
    \centering
\begin{tikzpicture}
  \begin{axis}[
     height = 4.5 cm,
     width = 0.99\linewidth,
     grid=both,
     thick,
     xmin = 0,
     xmax = 1,
     ymin = 4900,
     ymax = 5008,
     /pgf/number format/.cd,
     1000 sep={},
     xlabel={$\theta$},
     ylabel={Cost [CHF]},
     legend style={at={(0.2,0.55)},
		anchor=north}
    ]
 	\addplot+[matlabblue,thick,mark=o] coordinates {
	    (0.0,4995.6594)
	    (0.4,4929.0684)
	    (0.5,4912.0843)
   	    (0.6,4912.4588)
        (0.7,4947.7349)
        (0.8,4956.1573)
        (0.9,4963.7888)
		};	
	\addplot+[matlabyellow,thick,densely dotted,mark=o, mark options={solid}] coordinates {
	    (0.0,4995.6594)
	    (0.4,4975.4120)
	    (0.5,4920.5471)
	    (0.6,4925.4514)
        (0.7,4946.8236)
        (0.8,4964.3050)
        (0.9,4970.4607)
		};	
	\addplot+[matlabred,thick,dashed,mark=o, mark options={solid}] coordinates {
	    (0.0,4995.6594)
	    (0.4,4996.2644)
	    (0.5,5003.4239)
	    (0.6,4986.8554)
        (0.7,4969.7903)
        (0.8,4933.8852)
        (0.9,4946.3392)
		};
	\legend{$M_1 = 12$, $M_1 = 24$, $M_1 = 36$}
  \end{axis}
\end{tikzpicture}
    \caption{Objective of stochastic MPC algorithm, with decision a causal, saturated affine function of the initial realization $P_t^\mathrm{dem,net}$. Graphs depict effect of weighting initial $M_1$ hours of peak period by $\theta$ relative to entire peak, as in \eqref{eq:stochastic_mpc_subpeak_weighting}.  The simulations are run using a 4.5 day horizon, from March 1 - April 30, 2018.}
    \label{fig:stochastic_mpc_relative_subpeak_weighting}
\end{figure}

\subsubsection{Structure of decision policy}
We analyze the effect of improvements proposed in Section \ref{section:stochastic_problem} to the affine decision policy with saturation. As a baseline, the optimization problem of Section \ref{section:affine_decision_policy} with a minimum first timestep range as in \eqref{eq:net_demand_eq_mod} and \eqref{eq:first_timestep_noise} achieves an objective of 5101.10 CHF. When we include the intra-peak weighting scheme \eqref{eq:stochastic_mpc_subpeak_weighting} with $M_1 = 12$ and $\theta = 0.5$, this results in an objective of 4991.40 CHF. Adding the causal banded affine decision policy described in Section \ref{section:causal_timesteps} for all timesteps, with $M_2=48$, further improves the objective to 4912.08 CHF. 

\subsubsection{Minimum range of scenarios in first timestep}
In simulations of the various problem settings proposed in Section \ref{section:stochastic_problem}, it is beneficial to add a zero-mean Gaussian variable $\omega_t$ to the scenarios in the first timestep. Not including $\omega_t$ in the above case worsens the objective from 4912.08 to 4998.37. 

\subsubsection{Effect of relative weighting within peak}
The intra-peak weighting of \eqref{eq:stochastic_mpc_subpeak_weighting} improves the objective for the policy-based setting of Section  \ref{section:causal_stochastic_mpc}. We vary $\theta$ and $M_1$, and report the objectives in Fig.~\ref{fig:stochastic_mpc_relative_subpeak_weighting}. We see that certain choices of $\theta$ and $M_1$ improve the problem objective significantly compared to using the original $f^\mathrm{peak}(s_u,s_{u+1})$ (where $\theta = 0$). 

\subsubsection{Comparison between decision policies}
We choose the stochastic setting that performed best above; namely, adding a zero-mean Gaussian variable to the first timestep, using the intra-peak weighting \eqref{eq:stochastic_mpc_subpeak_weighting} with $M_1 = 12$ and $\theta = 0.5$, and fitting causal, banded affine decision policies for all timesteps, with $M_2 = 48$. We then compare various first timestep decision policies.  The first policy considered, where the action in the first timestep does not depend on the realization of the uncertainty, achieves an objective of 4906.21 CHF. The affine decision policy of \eqref{eq:causal_policy_structure} results in an objective of 4912.08 CHF as above, which is within 1.3\% of the optimum.

\subsection{Analysis of algorithm performance}
The policy with the decision for the first timestep fixed across all scenarios performed surprisingly well. This could occur because the conservatism of this policy counteracts the optimism of trusting the inaccurate forecasts. With more accurate forecasts, we expect an affine decision rule to perform better than such a conservative policy. 

The causal policy structure of Section \ref{section:causal_timesteps} is also beneficial, likely for several reasons. First, fitting policies that couple the decisions for each timestep across scenarios adds a degree of conservatism against overfitting. Second, the causal constraints emphasize near-term forecasts, which are more accurate. Timesteps closer to the present thus rely on more accurate portions of the forecast.

To put the performance of the chosen stochastic optimization problem formulation into perspective, we compare several simple operation strategies. Meeting the necessary demand from the grid directly without using storage results in a cost of 5711.30 CHF. Solving the one-shot deterministic optimization problem \eqref{eq:deterministic_problem} without the peak cost in the objective (i.e., using a modified objective of $\sum_{t=0}^{N-1} p_t^\mathrm{buy} P_t^\mathrm{grid} - p^\mathrm{term} E_N$), results in a solution cost, when evaluated using the true objective \eqref{eq:deterministic_objective}, of 5521.19 CHF. Finally, if we solve the deterministic optimization problem \eqref{eq:deterministic_problem}, but minimize the sum of the daily peaks as proposed in \cite{Hanna2014} (i.e., the term $\sum_{q=1}^Q s_q$ in \eqref{eq:deterministic_objective} consists of daily rather than monthly peaks $s_q$), the achieved objective is 5462.06 CHF.

\section{Conclusion}
We make many parameter and model choices in our approach to the peak shaving problem. With such a large design space, it is likely that other choices would lead to better results in other settings. However, the general techniques presented for weighting multiple peak periods relative to the energy cost, the direct fitting from meteorological  forecasts to PV and demand data, as well as the optimization over causal policies for the grid power, are all applicable to other settings. While the simulation results here are for one date interval, simulations run on other date intervals reveal similar relative performance between the methods analyzed. 

To increase the performance of our method for the objective considered, improving the forecasts would certainly help. For example, similar to the case of the PV forecast, a forecast model for the demand which uses simpler basis functions might be beneficial. We could also consider a more robust cost function for peak charge than simply the sum of costs over all scenarios. While the present paper only considers the given forecast horizon, an increased horizon length would allow the optimization problem to account for longer-term trends.

\section*{Acknowledgments}
The authors would like to thank the building control group at the Automatic Control Laboratory at ETH as well as the ESI platform group at PSI for insightful discussions that have improved this work. This work is supported by the Swiss Federal Office of Energy, under the project ReMaP.

%\addtolength{\textheight}{-4cm}   % This command serves to balance the column lengths



%\bibliographystyle{IEEEtran}

\begin{thebibliography}{17}

\bibitem{Glassmire2012}
J. Glassmire, P. Komor, and P. Lilienthal, ``Electricity demand savings from distributed solar photovoltaics,'' Energy Policy, vol. 51, pp. 323-331, 2012.

\bibitem{Neubauer2015}
J. Neubauer and M. Simpson, ``Deployment of Behind-The-Meter Energy Storage for Demand Charge Reduction.'' No. NREL/TP-5400-63162, National Renewable Energy Lab (NREL), Golden, CO (United States), 2015.

\bibitem{Ma2011}
J. Ma, J. Qin, T. Salsbury, and P. Xu, ``Demand reduction in building energy systems based on economic model predictive control,'' Chemical Engineering Science, vol. 67, no. 1, pp. 92-100, 2012.

\bibitem{Hanna2014}
R. Hanna, J. Kleissl, A. Nottrott, and M. Ferry, ``Energy dispatch schedule optimization for demand charge reduction using a photovoltaic-battery storage system with solar forecasting,'' Solar Energy, vol. 103, pp. 269-287, 2014.

\bibitem{Narimani2017}
M. R. Narimani, B. Asghari, and R. Sharma, ``Energy storage control methods for demand charge reduction and PV utilization improvement,'' 2017 IEEE PES Asia-Pacific Power and Energy Engineering Conference (APPEEC), Bangalore, 2017.

\bibitem{Conte2018}
F. Conte, S. Massucco, M. Saviozzi, and F. Silvestro, ``A Stochastic Optimization Method for Planning and Real-Time Control of Integrated PV-Storage Systems: Design and Experimental Validation,'' IEEE Transactions on Sustainable Energy, vol. 9, no. 3, pp. 1188-1197, 2018.

\bibitem{Yu2017}
J. Yu, J. Qin, and R. Rajagopal. ``On Certainty Equivalence of Demand Charge Reduction Using Storage.'' Proceedings of the American Control Conference 2017, pp. 3430–3437.

\bibitem{Jin2017}
J. Jin and Y. Xu, ``Optimal Storage Operation Under Demand Charge,'' IEEE Transactions on Power Systems, vol. 32, no. 1, pp. 795-808, Jan. 2017.

\bibitem{Park2017}
A. Park and P. Lappas, ``Evaluating demand charge reduction for commercial-scale solar PV coupled with battery storage,'' Renewable Energy, vol. 108, pp. 523-532, 2017.

\bibitem{Jones2017}
M. Jones and M. M. Peet, ``Solving dynamic programming with supremum terms in the objective and application to optimal battery scheduling for electricity consumers subject to demand charges,'' 2017 IEEE 56th Annual Conference on Decision and Control (CDC), Melbourne, VIC, pp. 1323-1329, 2017.

\bibitem{MeteoswissLink}
MeteoSwiss Federal Office of Meteorology and Climatology, ``COSMO-E – probabilistic forecasts for the Alpine region''
https://www.meteoswiss.admin.ch/home/measurement-and-forecasting-systems/warning-and-forecasting-systems/cosmo-forecasting-system.html, accessed 2018-10-08.

\bibitem{Chen2011}
C. Chen, S. Duan, T. Cai, and B. Liu, ``Online 24-h solar power forecasting based on weather type classification using artificial neural network.'' Solar Energy, vol. 85, pp. 2856–2870, 2011.

\bibitem{Reno2012} 
M.J. Reno, C.W. Hansen, and J.S. Stein, ``Global Horizontal Irradiance Clear Sky Models: Implementation and Analysis.'' Sandia Report, Mar 2012.

\bibitem{Dows1995}
R. Dows and E. Gough, ``PVUSA procurement, acceptance and rating practices for photovoltaic power plants.'' Pacific Gas and Electric Company, San Ramon, CA, Tech. Rep., 1995.

\bibitem{ADAMoptimizer}
D. P. Kingma and J. Ba, ``Adam: A method for stochastic optimization.'' arXiv preprint arXiv:1412.6980 (2014).

\bibitem{Lygeros2011}
D. Chatterjee, P. Hokayem, and J. Lygeros, ``Stochastic receding horizon control with bounded control inputs: a vector space approach,'' IEEE Transactions on Automatic Control, vol. 56, pp. 2704-2710, Nov 2011.

\bibitem{EKZ_price_legacy}
Elektrizitätswerke Kanton Zürich (EKZ), \textit{EKZ Tarife 2018}\\
https://polybox.ethz.ch/index.php/s/hZ8s8uuQmWSwdbY\\
(Archived version of https://www.ekz.ch/content/dam/ekz-internet/downloads/ekz-tarifsammlung.pdf, accessed 2018-10-08).


\end{thebibliography}

\end{document}